\documentclass[aps,pra,twocolumn,groupedaddress,showpacs,floatfix]{revtex4}
\usepackage{graphicx}

\begin{document}

\title{Chromatic dispersion of liquid crystal infiltrated capillary tubes and photonic crystal fibers}

\author{Per Dalgaard Rasmussen}\email{pdr@com.dtu.dk}
\author{Jesper L{\ae}gsgaard}
\author{Ole Bang}
\date{\today}

\address{\textbf{COM$\bullet$DTU} Department of Communications, Optics \& Materials, \\Technical University of Denmark,
{\O}rsteds Plads 345V, DK-2800 Kgs. Lyngby, Denmark}

\begin{abstract}
We consider chromatic dispersion of capillary tubes and photonic
crystal fibers infiltrated with liquid crystals. A perturbative
scheme for inclusion of material dispersion of both liquid crystal
and the surrounding waveguide material is derived. The method is
used to calculate the chromatic dispersion at different
temperatures.
\end{abstract}
\maketitle 

\section{Introduction}
Together with the development of photonic crystal fibers (PCFs), a
large amount of research has been devoted to investigate the
possibilities of infiltrating the air holes of a PCF with
different liquids\cite{kerbage2002optcom}, and thereby changing
the optical properties of the fiber. Depending on the refractive
index of the liquid, the guiding effect of the fiber can possibly
be changed from guiding based on modified total internal
reflection (mTIR), to guiding based on the photonic band gap
effect, where the core has a lower refractive index than the
effective index of the cladding. Also selective filling of PCFs,
where only some of the holes are infiltrated, has experienced a
considerable
interest\cite{nielsen2005joa,xiao2005opex,zografopoulos2006opex},
because this can be used to tailor the optical characteristics of
the PCF. Among the various liquids that can be infiltrated in a
PCF, liquid crystals (LC) distinguishes themselves, because of
their anisotropic nature, which allows the possibility of
controlling the optical parameters of the waveguide by changing
the orientation of the molecules\cite{larsen2003opex}. This
orientation can be controlled in different ways, for example by
applying an electric field externally. The optical characteristics
of the fiber can also be changed by varying the temperature, since
the ordinary and extraordinary refractive indices of LCs are
highly dependent on temperature. Recently these tunable properties
of LCs have been used in various experimental research
projects\cite{du2004apl,maune2004apl}.
\\The LC can be infiltrated in the PCF using various techniques, one
possibility is to use a pressure chamber, but this technique has
shown to introduce orientational irregularities in the alignment
of LC molecules\cite{phdthesisTanggaard}. Another possibility is
to infiltrate the holes of the PCF using capillary forces, this
technique has shown to give a regular alignment of the LC
molecules. A disadvantage using capillary forces for the
infiltration is that the length of the infiltrated region will
only be of the order of a few centimeters, while longer
infiltration lengths can be achieved using pressure infiltration.
\\In the present work we address the problem of calculating
chromatic dispersion curves for different waveguide designs, where
the material dispersion of both the LC and waveguide material is
taken into account. The LC infiltrated PCF structures we consider
have previously been studied
theoretically\cite{zografopoulos2006opex}, without inclusion of
material dispersion in the PCF material and LC, and only
considering the special case where the extraordinary index of the
LC and the index of the PCF material were identical. The material
dispersion of LC is important to take into account, since LCs are
highly dispersive, especially in the visible spectrum, where the
dispersion can be much stronger than in for example silica. It has
previously been shown that approximating the total dispersion
simply by adding the waveguide and material dispersion gives the
correct qualitative behavior of the dispersion
curve\cite{ferrando2000ol}, but is not sufficient if quantitative
data for the dispersion is needed, for example to determine the
position of zero dispersion wavelengths (ZDWs).\\
To calculate precise dispersion profiles, we must therefore
include the material dispersion in the field equations. This
destroys the well known scalability of Maxwells equations, hence
if all the physical dimensions of the fiber are multiplied by a
constant factor, we are not able to calculate the new dispersion
curve without having to solve the field equations again. In
addition, if the computational method used to find the eigenmodes
numerically takes the propagation constant as an input variable,
and returns the corresponding frequency, we must ensure that this
is done in a self-consistent manner, i.e. the values of the
dielectric constants in the numerical calculation must correspond
to the values of the dielectric constants at the frequency
returned by the computational method.\\
In this work we find the self-consistent frequencies based on a
generalization of a perturbative method developed for isotropic
waveguides\cite{laegsgaard2003josab}. We consider dispersion
profiles of both simple waveguides consisting of capillary silica
tubes infiltrated with LCs, and more advanced selectively filled
PCF structures. Finally we investigate how a change in temperature
affects the dispersion characteristics of the fiber.
\section{Theory}
\subsection{Alignments of LC-molecules}
\begin{figure}
  \centering
  \includegraphics[scale=0.2]{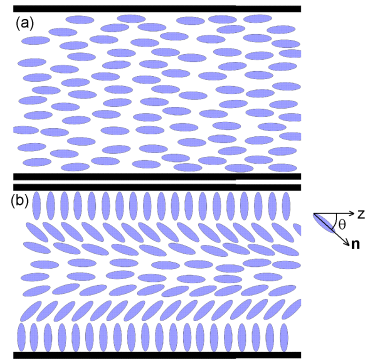}
  \caption{Parallel (a) and axial (b) alignment of LC molecules in a cylindrical geometry. Figures reproduced with permission\cite{phdthesisTanggaard}.}
  \label{ParAtanAlignFig}
  \end{figure}

In all the fiber designs considered in this work the LC is
contained inside a hollow circular cylinder. It is well known that
an intense optical field will interact with the LC and change the
orientation of the molecules\cite{tabiryan1986mclc}, here we
assume that the intensity of the optical field is so weak that we
can neglect this interaction. The LC is assumed to be in the
nematic phase, where the orientation of the molecules is
correlated, resulting in a preferred local orientation of the
molecules. This local orientation is described by the director
axis $\mathbf{n}(\mathbf{r})$ which is a unit vector pointing in
the same direction as the axis of the LC-molecules. In the general
case $\mathbf{n}(\mathbf{r})$ is found by minimization of elastic
energy. For the cylindrical geometry considered here the director
axis has the following form
$\mathbf{n}=(\sin(\theta),0,\cos(\theta))$ in cylindrical
coordinates $(r,\phi,z)$. $\theta$ is the angle between the
director axis $\mathbf{n}$ and the $z$-axis. The alignment of the
LC molecules in this situation has previously been studied
theoretically\cite{lin1991molcrystliqcryst}. $\theta(r)$ can be
found by solving a 2nd. order nonlinear ordinary differential
equation, where the only parameters are the elastic constants of
the LC. We assume that the molecules along $r=0$ are aligned
parallel to the $z$-axis ($\theta(0)=0$). The boundary condition
at the cylinder wall depends on the LC and the coating of the
capillary. Here we consider the two different possibilities
$\theta(R)=0$ and $\theta(R)=\pi/2$, where $R$ is the radius of
the cylinder. The dielectric tensor of a nematic LC is described
in terms of the perpendicular and parallel part of the optical
permittivity $\epsilon_{\bot}$ and $\epsilon_{||}$. If the
orientation of the molecules is described by the angle
$\theta(r)$, then the dielectric tensor has the following form in
cylindrical coordinates\cite{lin1991molcrystliqcryst}
\begin{eqnarray}\label{LCDielTensor}
\bar{\bar \epsilon}=\left(\begin{array}{ccc}
\epsilon_{rr}& 0 & \epsilon_{rz} \\
0 & \epsilon_{\phi\phi} & 0 \\
\epsilon_{zr} & 0 & \epsilon_{zz}
\end{array}\right),
\end{eqnarray}
where $\epsilon_{rr}=\epsilon_{\bot}+\Delta\epsilon\sin^2\theta$,
$\epsilon_{rz}=\epsilon_{zr}=\Delta\epsilon\sin\theta\cos\theta$,
$\epsilon_{\phi\phi}=\epsilon_{\bot}$ and
$\epsilon_{zz}=\epsilon_{\bot}+\Delta\epsilon\cos^2\theta$.
$\Delta\epsilon$ is the optical anisotropy defined as
$\Delta\epsilon=\epsilon_{||}-\epsilon_{\bot}$. If the 3 elastic
constants of the LC describing twist, splay and bend deformations
are assumed to be equal, and we further assume that the molecules
are aligned parallel with the $z$-axis in the center of the
cylinder, the orientation is given by
$\theta(r)=2\tan^{-1}(ar/R)$, where $a$ is a constant depending on
the boundary condition at the wall. For the boundary conditions
considered here we have the two simple analytical solutions
$\theta(r)=0$ and $\theta(r)=2\tan^{-1}(r/R)$, depending on
whether the molecules are anchored parallel or perpendicular to
the boundary of the cylinder. The two orientations are shown
schematically in Fig. \ref{ParAtanAlignFig}. In the following we
will consider these two orientations of the LC molecules, and
refer to them as planar ($\theta(r)=0$) and axial
($\theta(r)=2\tan^{-1}(r/R)$) alignment. The planar alignment is
easily achieved experimentally, while axial alignment requires
that the capillary is coated with a surfactant before the LC is
infiltrated\cite{phdthesisTanggaard}.
\subsection{Calculation of chromatic dispersion curves} In this
section we derive a perturbative method for calculation of
chromatic dispersion of a waveguide infiltrated with LC. The
method is general and can be applied to arbitrary waveguide
designs. Our method is a generalization of an earlier presented
method for isotropic waveguides\cite{laegsgaard2003josab}, but
this method allows the possibility that the waveguide consists of
anisotropic materials. We consider a waveguide which is uniform
along the $z$-direction, and therefore assume that the magnetic
field can be described in the form of a monochromatic wave
travelling along the $z$-direction, i.e.
$\mathbf{H}(x,y,z,t)=\exp[i(\beta z-\omega
t)]\mathbf{h}(x,y;\beta)$. From Maxwells equations the following
equation for the vector field $\mathbf{h}(x,y;\beta)$ is derived
\begin{eqnarray}\label{fundeq1}
\mathbf{\Phi}\mathbf{h}&=&\frac{\omega^2}{c^2}\mathbf{h},\\
\label{fundeq12}\mathbf{\Phi}&=&\nabla_{\beta}\times\bar{\bar{\epsilon}}^{-1}\nabla_{\beta}\times
\end{eqnarray}
where the operator $\nabla_{\beta}$ is given by
$\nabla_{\beta}=(\partial /\partial x,\partial /\partial
y,i\beta)$. $\bar{\bar{\epsilon}}$ is the dielectric tensor, which
in the LC region is given by the expression in Eq.
(\ref{LCDielTensor}). In the silica region the dielectric tensor
is simply a diagonal matrix, with the dielectric constant of
silica in the diagonal. Therefore
$\bar{\bar{\epsilon}}=\bar{\bar{\epsilon}}(\mathbf{r},\epsilon_S(\omega),\epsilon_\bot(\omega),\epsilon_{||}(\omega))$,
i.e. the dielectric function depends on position, the dielectric
constant of the material surrounding the LC ($\epsilon_S$), and
the dielectric constants $\epsilon_{\bot}$ and $\epsilon_{||}$ of
the LC. Since material dispersion is taken into account, all 3
dielectric constants are assumed to be frequency dependent. The
dispersion coefficient is defined by
\begin{equation}\label{dispersion}
D=\frac{\omega^2}{2\pi cv_g^2}\frac{d v_g}{d\omega},
\end{equation}
where $v_g$ is the group velocity, defined as
$v_g=\frac{d\omega}{d\beta}$. To find an exact expression for the
group velocity we start out by rewriting Eq. (\ref{fundeq1}) as
\begin{eqnarray}\label{fundeq1InnerProd}
\frac{\langle\mathbf{h},\mathbf{\Phi}\mathbf{h}\rangle}{\langle\mathbf{h},\mathbf{h}\rangle}=\frac{\omega^2}{c^2}.
\end{eqnarray}
Here and in the following we use the following notation for the
inner product $\langle\mathbf{A},\mathbf{B}\rangle=\int
\mathbf{A}^*\cdot \mathbf{B} d\mathbf{r}_{\bot}$, i.e. the
integration is over the whole transverse plane. Now the group
velocity is found by differentiating both sides of Eq.
(\ref{fundeq1InnerProd}) with respect to the propagation constant
$\beta$. The left hand side of Eq. (\ref{fundeq1InnerProd}) is
differentiated with respect to $\beta$ using the Hellman-Feynman
theorem. To differentiate the operator $\mathbf{\Phi}$ with
respect to $\beta$, we note that the operator depends on $\beta$
explicitly through $\nabla_{\beta}$, and implicitly through the
dielectric constants $\epsilon_{j}=\epsilon_{j}(\omega(\beta))$
($j=S,\bot,||$). In the following $\partial/\partial \beta$
denotes a differentiation for fixed dielectric constants. Using
this notation we have the following expression for $\mathbf{\Phi}$
differentiated with respect to $\beta$
\begin{eqnarray}\label{opediff}
\frac{d\mathbf{\Phi}}{d\beta}=\frac{\partial
\mathbf{\Phi}}{\partial \beta}+v_g\sum_j\frac{\partial
\mathbf{\Phi}}{\partial \epsilon_j}\frac{\partial
\epsilon_j}{\partial \omega}.
\end{eqnarray}
Where we have the following expressions for
$\partial\mathbf{\Phi}/\partial \beta$ and
$\partial\mathbf{\Phi}/\partial \epsilon_j$
\begin{eqnarray}
\frac{\partial
\mathbf{\Phi}}{\partial\beta}&=&\nabla_{\beta}\times\bar{\bar{\epsilon}}^{-1}\left(
\begin{array}{c}
0\\
0\\
i \end{array} \right)\times+\left(
\begin{array}{c}
0\\
0\\
i \end{array} \right)\times\bar{\bar{\epsilon}}^{-1}\nabla_{\beta}\times\\
\frac{\partial \mathbf{\Phi}}{\partial
\epsilon_j}&=&-\nabla_{\beta}\times\bar{\bar{\mathbf{\epsilon}}}^{-1}\frac{\partial\bar{\bar{\mathbf{\epsilon}}}
}{\partial
\epsilon_j}\bar{\bar{\mathbf{\epsilon}}}^{-1}\nabla_{\beta}\times
\end{eqnarray}
Using the Hermiticity of the operator $\nabla_\beta$, and the
Maxwell equation $\nabla_\beta \times
\mathbf{h}=-i(\omega/c)\bar{\bar{\epsilon}}\mathbf{e}$, now gives
us the general expression for the group velocity in the case where
the material dispersion $\epsilon_j=\epsilon_j(\omega)$ is known
\begin{eqnarray}
\label{groupvel}
v_g&=&v_g^0\left[1+\frac{\omega}{2}\left(\frac{\langle
\mathbf{e},\frac{d \epsilon_S}{d
\omega}\mathbf{e}\rangle_S}{\langle \mathbf{h},\mathbf{h}
\rangle}\right.\right.\\&&+\left.\left.
\frac{\langle\mathbf{e},\left(\frac{\partial\bar{\bar{\mathbf{\epsilon}}}
}{\partial
\epsilon_{\bot}}\frac{d\epsilon_{\bot}}{d\omega}+\frac{\partial\bar{\bar{\mathbf{\epsilon}}}
}{\partial
\epsilon_{||}}\frac{d\epsilon_{||}}{d\omega}\right)\mathbf{e}\rangle_{LC}}
{\langle \mathbf{h},\mathbf{h} \rangle} \right)\right]^{-1},
\end{eqnarray}
here $\langle\cdot,\cdot\rangle_{S}$ and
$\langle\cdot,\cdot\rangle_{LC}$ denote that the integration is
only over the silica or the LC respectively. The electric field
$\mathbf{e}$ is defined similarly to $\mathbf{h}$, i.e. it is the
part of the electric field where the $z$ and $t$ dependence has
been factored out. In Eq. (\ref{groupvel}) $v_g^0$ denotes the
group velocity when the material dispersion is zero. An exact
expression for $v_g^0$ is found by differentiation of Eq.
(\ref{fundeq1InnerProd}) with respect to $\beta$, and again using
the Hermiticity of $\nabla_\beta$ and the Maxwell equation
$\nabla_\beta \times
\mathbf{h}=-i(\omega/c)\bar{\bar{\epsilon}}\mathbf{e}$ i.e.
\begin{equation}\label{groupvelnoDisp}
v_g^0=\frac{c^2}{2\omega}\frac{\langle\mathbf{h},\frac{\partial
\mathbf{\Phi}}{\partial
\beta}\mathbf{h}\rangle}{\langle\mathbf{h},\mathbf{h}\rangle}=c\frac{Re\langle[\mathbf{e^*}\times\mathbf{h}]_z\rangle}{\langle\mathbf{h},\mathbf{h}\rangle}.
\end{equation}
where $\langle f\rangle=\int f d\mathbf{r}_{\bot}$. Our
perturbative method for calculating $D$ consists of several steps,
first we make a guess for self-consistent values of the dielectric
constants $\epsilon_{i,0}$ and solve Eq. (\ref{fundeq1}). From
this solution we find the nonselfconsistent frequency $\omega_0$,
and the group velocity due to waveguide dispersion $v_g^0$ by
using the definition in Eq. (\ref{groupvelnoDisp}). Now
generalizing the procedure for isotropic waveguides, we see that a
first order approximation to the self-consistent frequency is
\begin{equation}
\label{omegasc}\omega_{sc}\approx \omega_0+\sum_j\frac{\partial
\omega}{\partial
\epsilon_j}\Delta\epsilon_j=\omega_0\left(1-\sum_jE_j\Delta
\epsilon_j\right),
\end{equation}
where $E_j$ ($j=S,\bot,||$) is given by
\begin{eqnarray}
E_S&=&\frac{1}{2}\frac{\langle\mathbf{e},\mathbf{e}\rangle_{S}}{\langle\mathbf{h},\mathbf{h}\rangle},\\
E_{\bot}&=&\frac{1}{2}\frac{\langle\mathbf{e},\frac{\partial \bar{\bar{\mathbf{\epsilon}}}}{\partial \epsilon_{\bot}}\mathbf{e}\rangle_{LC}}{\langle\mathbf{h},\mathbf{h}\rangle},\\
E_{||}&=&\frac{1}{2}\frac{\langle\mathbf{e},\frac{\partial
\bar{\bar{\mathbf{\epsilon}}}}{\partial
\epsilon_{||}}\mathbf{e}\rangle_{LC}}{\langle\mathbf{h},\mathbf{h}\rangle}.
\end{eqnarray}
In Eq. (\ref{omegasc}) $\partial \omega /\partial
\epsilon_j=-\omega E_j$ has been found by differentiating Eq.
(\ref{fundeq1InnerProd}) with respect to $\epsilon_j$.
$\Delta\epsilon_j$ in Eq. (\ref{omegasc}) is found by noting that
\begin{eqnarray}\label{linsysDeps}
\epsilon_i(\omega_{sc})&\approx&
\epsilon_i(\omega_0)+\frac{d\epsilon_i}{d\omega}|_{\omega_0}(\omega_{sc}-\omega_0)\\&=&\epsilon_{i}(\omega_{0})-\frac{d
\epsilon_i}{d
\omega}|_{\omega_0}\sum_{j}E_j\Delta\epsilon_{j}\omega_0\\&=&\epsilon_{i,0}+\Delta\epsilon_i.
\end{eqnarray}
Eqs. (\ref{linsysDeps}) define a system of 3 coupled linear
algebraic equations. Once Eqs. (\ref{linsysDeps}) have been solved
for $\Delta\epsilon_i$, our approximation to the selfconsistent
frequency is readily found using Eq. (\ref{omegasc}). Since our
goal is to use Eq. (\ref{groupvel}) for finding the group
velocity, we must also find an approximation to $v_g^0$. This is
done by using that
\begin{equation}
\frac{\partial v_g^0}{\partial
\epsilon_j}=\frac{\partial^2\omega}{\partial \epsilon_j
\partial \beta}=-\frac{\partial}{\partial \beta}\omega E_j=-(v_g^0E_j+\omega\frac{\partial E_j}{\partial \beta}).
\end{equation}
A first order approximation to the self-consistent group velocity
is then found using Eq. (\ref{groupvel})
\begin{equation}\label{selfconsisgroupvel}
v_g^{sc}=\frac{v_g^0-\sum_j(v_g^0E_j+\omega_0\frac{\partial
E_j}{\partial
\beta})\Delta\epsilon_j}{1+\omega_{sc}\sum_jE_j\frac{d
\epsilon_j}{d \omega}|_{\omega_{sc}}},
\end{equation}
In Eq. (\ref{selfconsisgroupvel}) and (\ref{linsysDeps}) we find
the derivatives of the dielectric constants by differentiating the
Sellmeier or Cauchy polynomial presented in the following with
respect to frequency. $E_j$ is found from the fields returned by
the computational method, when using the dielectric constants
$\epsilon_{i,0}$. Strictly speaking the values of $E_j$ used in
Eq. (\ref{selfconsisgroupvel}) should be the values calculated at
the selfconsistent frequency. This would require solving Eq.
(\ref{fundeq1}) more than one time, and therefore significantly
increase the calculation time for each propagation constant, but
we have found that the variations in $E_j$ with $\epsilon_i$ can
safely be neglected. The derivatives of $E_j$ with respect to
$\beta$ are found using a standard three point approximation.
After having found the selfconsistent frequencies and the
corresponding group velocities for a number of propagation
constants $\beta$, we find the dispersion $D$ by using the
definition given in Eq. (\ref{dispersion}), in this calculation
the derivative of the group velocity $v_g$ with respect to
frequency $\omega$ is also approximated using a three point
formula.

\section{Results}
Both the planar and axial alignment discussed in the previous
section are considered. For the material dispersion of silica we
use the Sellmeier curve
\begin{equation}\label{selmeierEq}
\epsilon_{SiO_2}=1+\sum_{j=1}^{3}\frac{a_j\lambda^2}{\lambda^2-b_j},
\end{equation}
where $a_i$ and $b_i$ are constants. Here we use the values in
Table \ref{SellmeierCauchyParms} as reported by
Okamoto\cite{okamoto}.
\begin{table}
\centering
\begin{tabular}{l l l l l l l}
$i$ &   $A_i^{\bot} (25^o$C$)$    &  $A_i^{||}  (25^o$C$)$&  $A_i^{\bot}  (50^o$C$)$&  $A_i^{||}  (50^o$C$)$  &  $a_i$ & $b_i$  \\\hline \hline\\
$1$ & $1.4994$ & $1.6933$ &1.5062&1.6395& $0.6965325$ &
$4.368309\cdot
10^{-3}$\\
$2$ & $0.0070$ & $0.0078$ &0.0063&0.0095&$0.4083099$ &
$1.394999\cdot
10^{-2}$\\
$3$ & $0.0004$ & $0.0028$ &0.0006&0.0020& $0.8968766$ & $97.93399$\\
 \hline \hline
\end{tabular}

\caption{Parameters for Cauchy and Sellmeier polynomials given in
Eq. (\ref{selmeierEq}-\ref{cauchyEq}). The non-dimensionless
parameters are all given in units of $\mu m^2$ or $\mu m^4$.}
\label{SellmeierCauchyParms}
\end{table}

\begin{figure}
  \centering
  \includegraphics[scale=0.2]{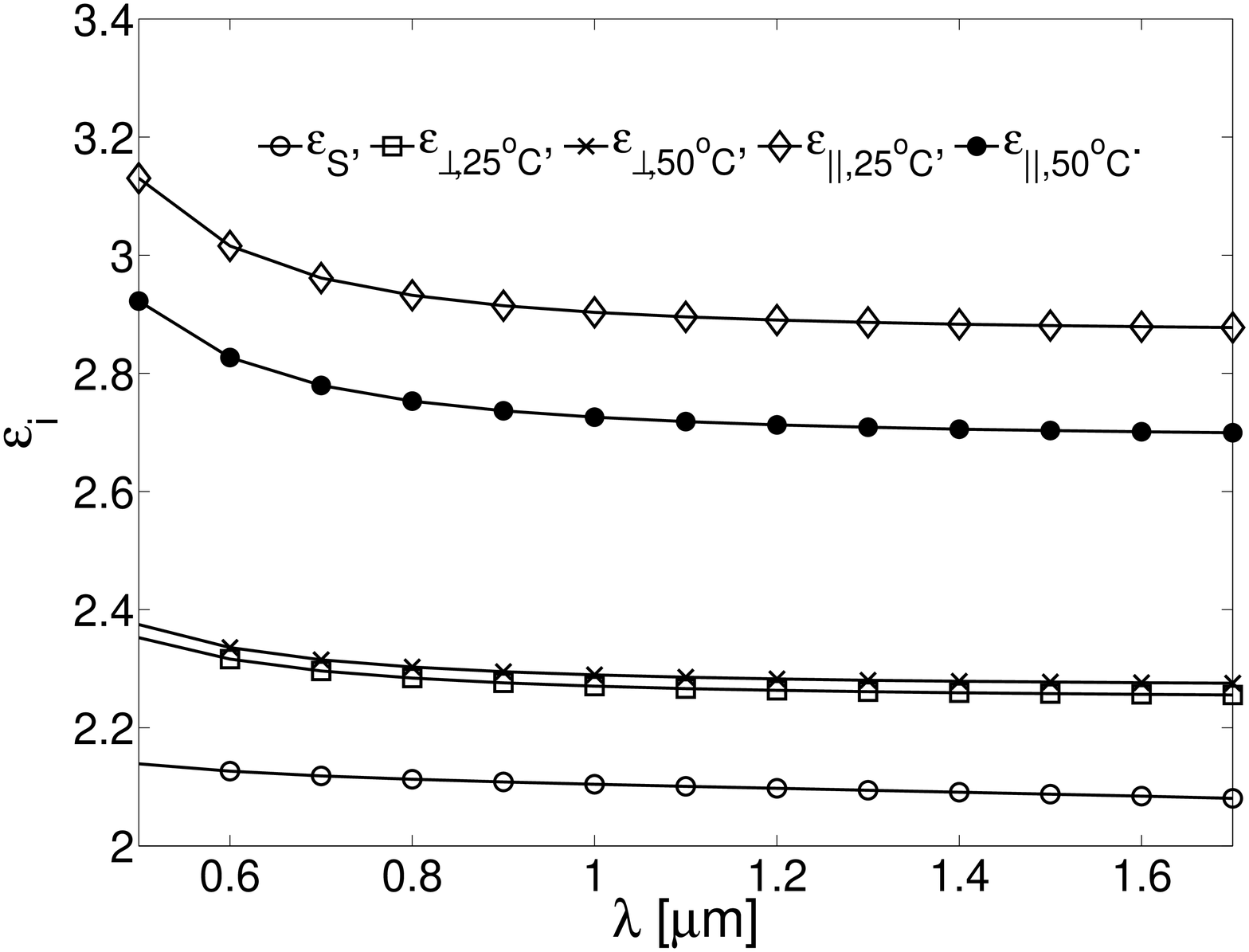}
  \caption{Dielectric constants of silica and E7. The curves for silica and
           E7 are based on the Sellmeier expression given in Eq. (\ref{selmeierEq}) and
           the Cauchy polynomial given in Eq. (\ref{cauchyEq}) respectively. The
           parameters given in Table \ref{SellmeierCauchyParms} are used.}
  \label{SellmeierCauchyPlots}
  \end{figure}

For the LC a Cauchy polynomial is used for both $\epsilon_{\bot}$
and $\epsilon_{||}$
\begin{equation}\label{cauchyEq}
\epsilon_{\bot,||}=\left(A_1^{\bot,||}+\frac{A_2^{\bot,||}}{\lambda^2}+\frac{A_3^{\bot,||}}{\lambda^4}\right)^2
\end{equation}
where $A_i^{\bot,||}$ are constants. Here we use the values for
the liquid crystal E7 given in Table \ref{SellmeierCauchyParms} as
reported by Li \emph{et al}\cite{li2005jap} at $25^o$C and
$50^o$C. The values obtained for the dielectric constants using
these parameters in the Cauchy polynomials have been shown to be
consistent with measured values throughout the visible spectrum
and far into the infrared spectrum. In Fig.
\ref{SellmeierCauchyPlots} we have plotted the three dielectric
constants $\epsilon_S$, $\epsilon_{\bot}$ and $\epsilon_{||}$ as a
function of vacuum wavelength.
We see that the dielectric constant of silica ($\epsilon_S$) is
below the two dielectric constants of E7 ($\epsilon_{\bot}$ and
$\epsilon_{||}$) throughout the visible spectrum and into the near
infrared spectrum, hence a waveguide based on TIR can be realized
in this spectrum. In the following we examine the chromatic
dispersion for different waveguides based on TIR for the capillary
tubes, and modified TIR for the PCFs. Whenever the perturbative
method is used to obtain self-consistent frequencies, the guesses
for the self-consistent values of the dielectric constants
$\epsilon_{j,0}$ are taken to be the values corresponding to a
vacuum wavelength of $1\mu$m. We solve Eq. (\ref{fundeq1}) using a
freely available software package\cite{johnson2001:mpb} where the
electric field is expanded in plane waves. In this software
package periodic boundary conditions are assumed on all
boundaries, therefore all calculations for both the single
capillary and the PCF structure are done using a supercell which
is considerably larger than the LC infiltrated cylinder in order
to minimize interactions between the images. In this work the
distance between repeated images was 14 relative to the radius of
the LC infiltrated cylinder for the capillary tubes, and 14
relative to the pitch for the PCF structures. Each elementary cell
of the supercell consisted of a uniform $32\times 32$ grid. The
relative error using these parameters was estimated to be below
$5\%$, by repeating a set of the computations on a finer $64\times
64$ grid.
\subsection{Capillary tube infiltrated with LC}
\begin{figure}
  \centering
  \includegraphics[scale=0.2]{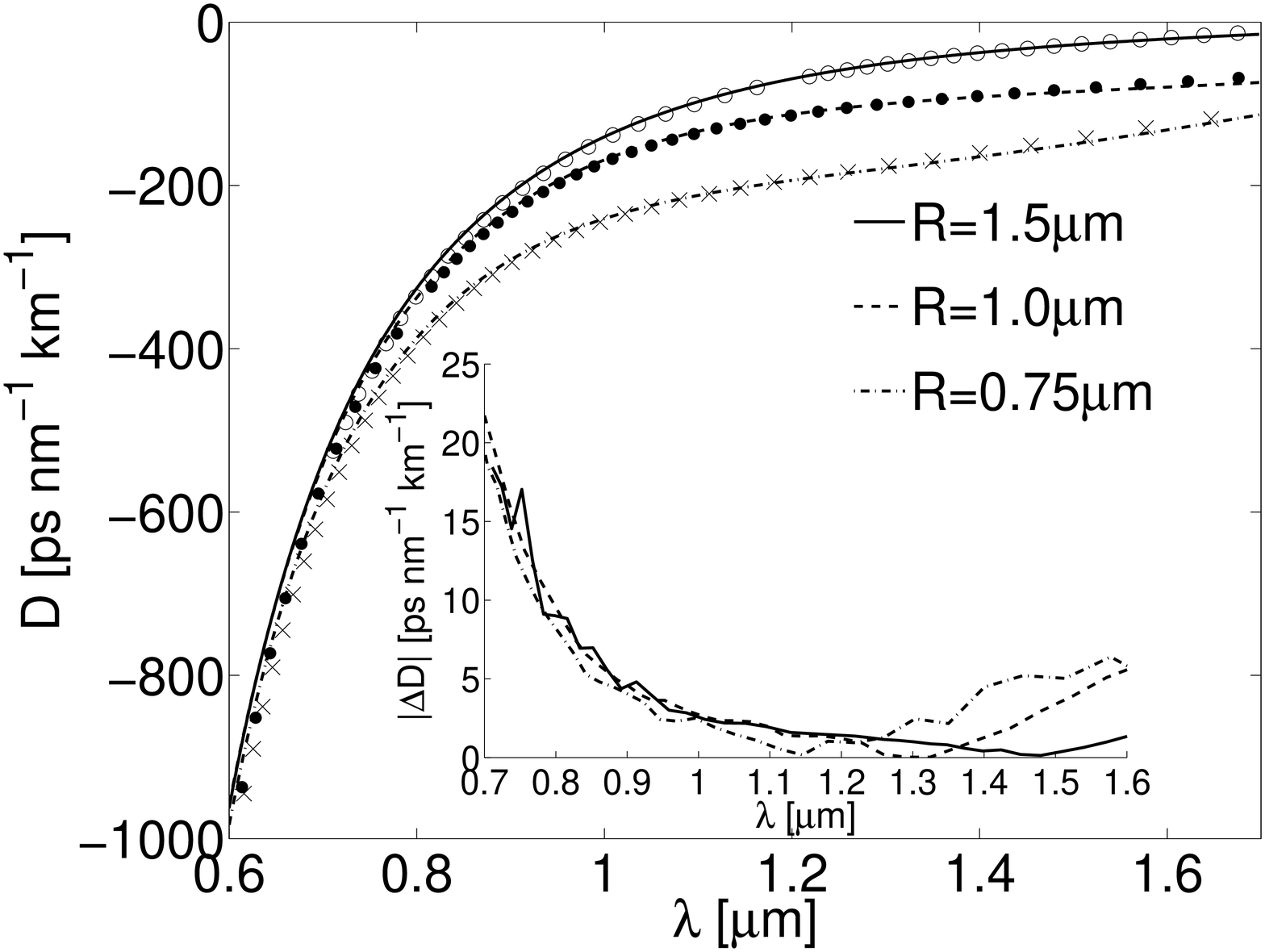}
  \caption{Chromatic dispersion of the fundamental HE$_{11}$ mode for planar alignment of LC molecules and different radii
           of the capillary tube. The lines show the chromatic dispersion found using the
           exact analytical result for the dispersion relation, and markers show the result
           found numerically together with our perturbative method. The inset shows the
           absolute difference between the two results, i.e. $|\Delta D|=|D_{exact}-D_{perturbative}|$.}
  \label{AnalyticalDispPerturDisp}
  \end{figure}

\begin{figure}
  \centering
  \includegraphics[scale=0.2]{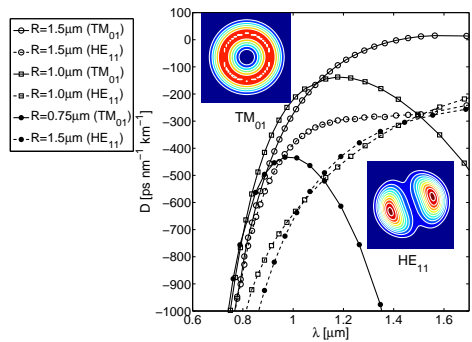}
  \caption{Chromatic dispersion for axial alignment of LC molecules and different radii
           of the capillary tube. The self-consistent frequencies are found using the
           perturbative method. Examples of $|\mathbf{H}|^2$ for the TM$_{01}$ and HE$_{11}$ mode are shown on the insets.}
  \label{PerturDispAtan}
  \end{figure}

\begin{figure}
  \centering
  \includegraphics[scale=0.2]{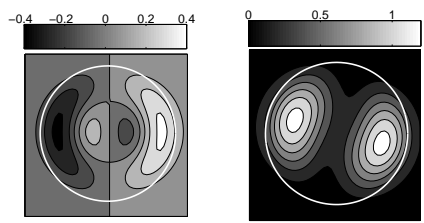}
  \caption{Representative contour plots of real part of the $x$-component of $\mathbf{H}$ for the TM$_{01}$ mode (left) and the HE$_{11}$ mode (right) for the axial alignment in a capillary tube. The white circle shows the boundary of the LC infiltrated capillary.}
  \label{oddevenfig4}
  \end{figure}

\begin{figure}
  \centering
  \includegraphics[scale=0.2]{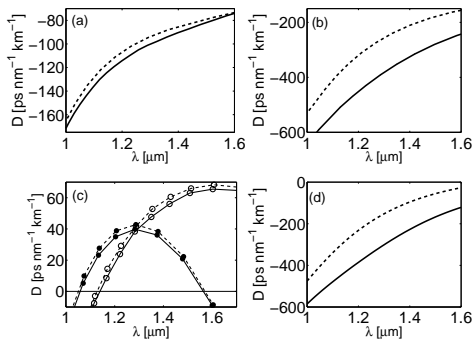}
  \caption{Chromatic dispersion at $25^o$C (solid lines) and at $50^o$C (dashed lines) for HE$_{11}$ modes.
           (a)-(b) Capillary tube ($R=1.0\mu$m) with LC parallel and axially aligned respectively. (c)-(d)
           Selectively filled PCF structure as shown in inset in Fig. \ref{PerturDispPCFplanar} with LC parallel and
           axially aligned respectively. In subplot (c) the bullets ($\bullet$) and circles ($\circ$) are for
           $R=0.75\mu$m and $R=1.0\mu$m respectively. The dispersion profile in (d) is for $R=1.0\mu$m.}
  \label{Temp50C}
  \end{figure}

First we consider a simple waveguide consisting of a circular hole
containing LC surrounded by a silica cladding. Such a waveguide
can be realized physically by infiltrating a capillary tube with
LC. If we assume that the LC molecules align in the planar
orientation discussed in the previous section, the dielectric
tensor in Eq. (\ref{LCDielTensor}) only has nonzero elements in
the diagonal, i.e. $\bar{\bar
\epsilon}=diag(\epsilon_{\bot},\epsilon_{\bot},\epsilon_{||})$. If
we further assume the cladding has infinite width, an analytical
solution to Eq. (\ref{fundeq1}) can be derived\cite{dai1991josaa}.
We can therefore use this solution to investigate the accuracy of
our numerical perturbative method. The fundamental mode of the
waveguide considered here is always the HE$_{11}$ mode. For a
certain mode to be guided in this structure, the propagation
constant $\beta$ must satisfy
$\epsilon_{S}^{1/2}k<\beta<\epsilon_{\bot}^{1/2}k$, where $k$ is
the vacuum wavenumber $k=\omega/c$. The fiber has a single guided
mode when the $V$-parameter
($V=kr(\epsilon_{\bot}-\epsilon_{SiO_2})^{1/2}$) is less than
$2.405$, where $r$ is the inner radius of the tube. In the
following we consider fibers with radii of $1.5\mu$m, $1.0\mu$m
and $0.75\mu$m, these fibers are single mode for wavelengths
larger than $1.63\mu$m, $1.06\mu$m and $0.81\mu$m respectively. In
Fig. \ref{AnalyticalDispPerturDisp} we have compared the chromatic
dispersion found analytically with the chromatic dispersion found
using the numerical method described above, together with the
perturbative method described in the theory section.
We see that the dispersion curves found numerically together with
the perturbative method are quantitatively consistent with the
exact dispersion curves. In the inset in Fig.
\ref{AnalyticalDispPerturDisp} the difference between the
analytical result and the perturbative result is also plotted, we
see that the smallest deviations between the two results occur in
the infrared region, this is also expected since the material
dispersion is lowest in this region (see Fig.
\ref{SellmeierCauchyPlots}). The relative error of the
perturbative method is below $5$\% in the wavelength interval from
$0.6\mu$m to $1.7\mu$m. The dispersion is very high in the visible
spectrum, which is mainly due to the high material dispersion in
this region. Also notice that the dispersion is normal ($D<0$) for
all the wavelengths and radii
considered for the planar alignment.\\
For the axial alignment of the LC molecules there only exists an
analytical solution to Eq. (\ref{fundeq1}) for the TE
modes\cite{lin1991molcrystliqcryst}. But since the fundamental
mode, i.e. the mode with the lowest frequency, is not a TE-mode we
must solve Eq. (\ref{fundeq1}) numerically to find the chromatic
dispersion for the fundamental mode. It turns out that at short
wavelengths the fundamental mode is the TM$_{01}$ mode, while for
longer wavelengths the fundamental mode is the hybrid HE$_{11}$
mode. For the tube radii and wavelengths considered here, the
capillary tubes with the axial orientation are always multimoded.
Here we consider the two modes with lowest frequency; the
HE$_{11}$ and TM$_{01}$ mode. The dispersion of these modes as a
function of vacuum wavelength is shown in Fig.
\ref{PerturDispAtan}.
Again we see that the dispersion is mostly normal for the
wavelengths and tube radii considered here. But for $r=1.5\mu$m
the dispersion becomes anomalous for vacuum wavelengths higher
than approximately $1.4\mu$m for the TM$_{01}$ mode. For
$r=1.0\mu$m the fundamental mode switches from the HE$_{11}$ mode
to the TE$_{01}$ mode at a vacuum wavelength around
$\lambda=1\mu$m. For $r=0.75\mu$m and $r=1.5\mu$m the switch
between the two modes happens below $\lambda=0.6\mu$m and above
$\lambda=1.7\mu$m respectively. In an experimental setup light is
coupled into the LC infiltrated region using the HE$_{11}$ mode of
a single mode step index fiber which is an even mode. Therefore it
will most likely be easiest to excite the HE$_{11}$ mode of the LC
infiltrated region, since this mode is also even, in contrast to
the TM$_{01}$ which is odd. This is demonstrated in Fig.
\ref{oddevenfig4}, where the real part of the $x$-component of
$\mathbf{H}$ is plotted for the TM$_{01}$ and HE$_{11}$.
A similar behavior is found for the other components of the
$\mathbf{H}$-field, hence the TM$_{01}$ mode is odd (even though
the intensity plot of $|\mathbf{H}|^2$ in Fig.
\ref{PerturDispAtan} is even), and the
HE$_{11}$ mode is even.\\
In the following the effect of increasing the temperature to
$50^o$C will be studied. We do not consider temperatures above
$50^o$C, since the clearing temperature, i.e. the temperature
where $\epsilon_{\bot}=\epsilon_{||}$, is around $57^o$C for
E7\cite{li2005jap}. Above the clearing temperature the LC is no
longer in the anisotropic nematic phase. The parameters for the
ordinary and extraordinary indices of refraction at $50^o$C can
also be found with Cauchy polynomials. The coefficients at $50^o$C
are given in Table \ref{SellmeierCauchyParms}, and the dielectric
constants at $50^o$C are plotted in Fig.
\ref{SellmeierCauchyPlots}. We see that the increase in
temperature also increases the ordinary dielectric constant, while
the extraordinary dielectric constant is lowered. In subplot (a)
and (b) in Fig. \ref{Temp50C} the effect of raising the
temperature to 50$^o$C is shown for the capillary tube with the
planar and axial alignment of the LC molecules.
We see that the dispersion increases for both alignments. For the
planar alignment the explanation for this is straight forward. The
HE$_{11}$ mode carries most of its energy in the transverse
components of the field, and since the transverse components
experiences the ordinary dielectric constant, which increases with
temperature, the temperature increase effectively increases the
index difference between the core and cladding. The increased
index difference gives rise to the higher dispersion. For the
axial alignment the explanation for the increased dispersion is
more complicated than for the planar alignment. Here a field which
is mostly transverse will experience $\epsilon_{\bot}$ near the
center of the cylinder, and $\epsilon_{||}$ near the wall of the
cylinder. Since $\epsilon_{\bot}$ increases and $\epsilon_{||}$
decreases with temperature, as shown in Fig.
\ref{SellmeierCauchyPlots}, it is difficult to say a priori
whether the dispersion is increased or decreased.
\subsection{PCF infiltrated with LC}
\begin{figure}
  \centering
  \includegraphics[scale=0.2]{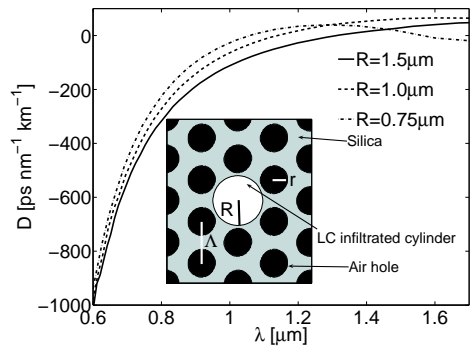}
  \caption{Chromatic dispersion for the PCF structure shown in the inset with different
            radii of the central hole. The ratios $R/\Lambda=0.6$ and $r/\Lambda=0.3$ are fixed. The LC molecules are planarly aligned. A zoom around the ZDWs is shown in Fig. \ref{Temp50C}.}
  \label{PerturDispPCFplanar}
  \end{figure}

In this section we consider a PCF-design similar to the structure
recently investigated by Zografopoulos \emph{et
al}\cite{zografopoulos2006opex}. Where the possibility of changing
the fiber characteristics by applying an external electric field
was considered. The structure has a cladding consisting of
airholes placed in a triangular structure and a core hole
infiltrated with LC. A cross section of the considered structure
is shown in the inset in Fig. \ref{PerturDispPCFplanar}.
  \begin{figure}
  \centering
  \includegraphics[scale=0.2]{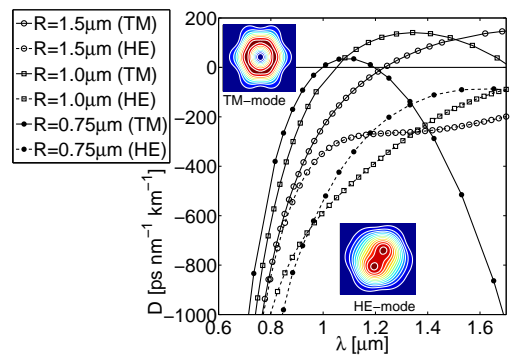}
  \caption{Chromatic dispersion for the PCF structure shown in inset in Fig. \ref{PerturDispPCFplanar} with different
            radii of the central hole. The LC molecules are aligned axially. Examples of $|\mathbf{H}|^2$ for the TM and HE mode are shown on the insets.}
  \label{PerturDispAtanPCF}
  \end{figure}
A physical realization of such a structure will require selective
filling, which has recently been
demonstrated\cite{nielsen2005joa,xiao2005opex}, where selective
filling was achieved by collapsing the small holes using a fusion
splicer, and then the holes with the larger radius were
infiltrated. Here we consider a structure where the infiltrated
center hole has a radius twice as large as the radius of the
cladding holes. The pitch is $5/3$ times the radius of the central
hole. Compared to the capillary tube studied in the previous
section this structure has a higher index contrast between the
core and cladding, because the presence of the airholes
significantly lowers the effective index of the cladding. Again we
consider both the planar and axial orientation of the LC in the
center hole. In Fig. \ref{PerturDispPCFplanar} the dispersion
curves for different radii of the center hole are shown for the
fundamental mode HE-mode with the planar alignment of the LC. The
fiber is multimoded for the wavelengths considered here. We see
that all the fibers now have regions of both normal and anomalous
dispersion. The fiber with a center hole radius of $0.75\mu$m has
two ZDWs at $\lambda=1.05\mu$m and $\lambda=1.55\mu$m. The fibers
with center hole radii of $1.0\mu$m and $1.5\mu$m each have one
ZDW at $\lambda=1.15\mu$m and $\lambda=1.3\mu$m respectively. The
dispersion curves for the axial alignment of the LC are shown in
Fig. \ref{PerturDispAtanPCF}. Like the capillary tube with the
axial alignment, the type of the fundamental mode is also
dependent on the wavelength for the PCF with the LC axially
aligned. We see that the fibers with center hole radius $1.0\mu$m
and $1.5\mu$m, now have large regions where the dispersion is
anomalous for the mode that resembles the TM$_{01}$ mode of the
single capillary tube. For the mode that resembles the
HE$_{11}$-mode of the single capillary tube the dispersion is
purely normal for all the waveguide designs considered here.\\
The effect of increasing the temperature is also investigated for
this waveguide design. In plots (c) and (d) in Fig. \ref{Temp50C},
we see again that the dispersion increases with temperature. In
subplot (c) the dispersion profiles for the PCFs with $R=0.75\mu$m
and $R=1.0\mu$m are shown. The lower ZDW of the fiber with
$R=0.75\mu$m can be tuned between $1.051\mu$m at $25^o$C and
$1.058\mu$m at $50^o$C, while the higher ZDW can be tuned between
$1.565\mu$m and $1.571\mu$m. The fiber with $R=1.0\mu$m has one
ZDW in the optical spectrum which can be tuned between $1.128\mu$m
and $1.139\mu$m. For the PCFs with the axial alignment of the LC
molecules we do not have anomalous dispersion for the designs
considered here. But the plots for for the axial alignment in Fig.
\ref{Temp50C} indicates that the dispersion can be tuned in a
broader interval for this alignment.
\section{Conclusion} An accurate method for
calculating chromatic dispersion of anisotropic waveguides is
demonstrated. The method is based on a generalization of a
previously presented method for isotropic waveguides. We have
applied the method to a simple step index fiber with an
anisotropic LC core since this problem has an analytical solution.
Our results show that the method can be applied to calculate
chromatic dispersion curves that are consistent with the exact
result throughout the visible spectrum and into the near infrared
spectrum.
\\With the method we have studied chromatic dispersion
of capillary tubes and PCFs infiltrated with LC. The considered
PCFs are all multimoded in the wavelength intervals considered,
while it is shown that single mode operation is possible for the
capillary tube infiltrated with LC molecules aligned in parallel.
The tunability of the different LC infiltrated waveguides is
investigated by calculating the chromatic dispersion at $25^o$C
and at $50^o$C. For the two different alignments of the LC
considered here, the tunability is highest for the axial
orientation, while the tunability for the planar orientation is
weaker. A waveguide design where the ZDWs can be tuned over
approximately $10\mu$m is demonstrated.



\newpage
\end{document}